\documentclass[sn-mathphys]{sn-jnl}

\usepackage{epsfig}
\usepackage{graphicx}
\usepackage{amsmath}
\usepackage{amssymb}
\usepackage{subfigure}
\usepackage{booktabs}
\usepackage{etoolbox}
\usepackage{setspace}

\jyear{2021}%

\theoremstyle{thmstyleone}%
\theoremstyle{thmstyletwo}%

\theoremstyle{thmstylethree}%

\begin{document}

\title[Deep Learning for Breast MRI Style Transfer with Limited Training Data]{Deep Learning for Breast MRI Style Transfer with Limited Training Data}


\author[1]{\fnm{Shixing} \sur{Cao}}

\author*[1]{\fnm{Nicholas} \sur{Konz {\small(ORCID: 0000-0003-0230-1598)}}}

\author[2]{\fnm{James} \sur{Duncan {\small(ORCID: 0000-0002-5167-9856)}}}

\author[1,3]{\fnm{Maciej A.} \sur{Mazurowski {\small(ORCID: 0000-0003-4202-8602)}}}

\affil[1]{\orgdiv{Department of Electrical and Computer Engineering}, \orgname{Duke University}, \orgaddress{\city{Durham}, \postcode{27704}, \state{NC}, \country{USA}}}

\affil[2]{\orgdiv{Department of Biomedical Engineering}, \orgname{Yale University}, \orgaddress{\city{New Haven}, \postcode{06520}, \state{CT}, \country{USA}}}

\affil[3]{\orgdiv{Departments of Computer Science, Radiology and Biostatistics \& Bioinformatics}, \orgname{Duke University}, \orgaddress{\city{Durham}, \postcode{27704}, \state{NC}, \country{USA}}}




\maketitle

\section*{Abstract}{\textbf{Purpose:}  In this work we introduce a novel medical image style transfer method, StyleMapper, that can transfer medical scans to an unseen style with access to limited training data. This is made possible by training our model on unlimited possibilities of simulated random medical imaging styles on the training set, making our work more computationally efficient when compared with other style transfer methods. Moreover, our method enables arbitrary style transfer: transferring images to styles unseen in training. This is useful for medical imaging, where images are acquired using different protocols and different scanner models, resulting in a variety of styles that data may need to be transferred between. \textbf{Methods:} Our model disentangles image content from style and can modify an image's style by simply replacing the style encoding with one extracted from a single image of the target style, with no additional optimization required. This also allows the model to distinguish between different styles of images, including among those that were unseen in training. We propose a formal description of the proposed model. \textbf{Results:} Experimental results on breast magnetic resonance images indicate the effectiveness of our method for style transfer. \textbf{Conclusion:} Our style transfer method allows for the alignment of medical images taken with different scanners into a single unified style dataset, allowing for the training of other downstream tasks on such a dataset for tasks such as classification, object detection and others.}

\section*{Introduction}
The same object can be depicted in an image in different styles. For example, a building can be shown in a photograph, a painting by a specific artist, or a sketch. Within the field of medical imaging, different styles manifest as data obtained by different scanner models and/or manufacturers. Deep learning, a subfield of artificial intelligence based on \textit{artificial neural networks}, has demonstrated an exceptional ability of solving image analysis problems. However, such a difference in style can be detrimental to these methods because it violates their common assumption that training and testing data possess the same style \cite{ma2019neural}. Style transfer methods, such as the one introduced in this paper, were proposed to address this problem in deep learning. 

Style transfer is a methodology that aims to preserve the consistency of the content of an image while changing the visual ``style''. Building upon this, \textit{Arbitrary} style transfer aims to transfer images to new styles \textit{unseen} in training, during which the content of the image can be transferred to the new style with minimal or zero additional model optimization. Preserving content is crucial in the medical imaging field because it is very important to ensure that underlying anatomical structure is preserved throughout the transformation process, and changing it could negatively impact the accuracy of diagnosis.

This task of unseen style transfer is important to develop for use within the medical setting. As an example: consider the case of one hospital, Hospital A, having MRI data of one style, Style A (e.g. GE scanner). Now, Hospital A receives certain data from another Hospital, Hospital B, of unknown or unseen style, Style B (e.g. Siemens scanner). If Hospital A wishes to use a model trained at Hospital B on images of Style B, but on their own Style A data, even if Hospital B only provides one or two images of Style B to Hospital A, our model could be used to extract the style code of Style B, and transfer all of Hospital A's Style A data to Style B, allowing Hospital A to use Hospital B's model on its own data. Section \ref{sec:exp:classifyscanners} involves an experiment that explores this exact scenario, where our model, StyleMapper, is used to transfer images of one MRI style to another MRI style unseen in training.

At a high level, our method learns to extract informative disentangled numerical representations of style and anatomical content of images. These representations, or style and content \textit{codes}, are obtained by inputting an image to a trained style encoder and content encoder. A pair of style and content codes, possibly from different images, can then be combined via a \textit{decoder} to synthesize a new image that contains the encoded content, but in the style described by the style code; both the encoders and the decoder are neural networks.

We introduce our model beginning with Section \ref{sec:basemodel}, which introduces our method of training our model to extract style and content codes from both raw image data and images transferred to simulated styles. The simulated style images are created by applying randomly sampled image transformation functions to raw images; these transformations are well-representative of the range of many styles/scanner types seen within medical imaging. In this way, the style transfer model sees a different style at each iteration of training. Because the image transformation functions have continuously-random parameters, the model can observe practically unlimited distinct styles during training, giving the style encoder more styles to learn from. This characteristic allows the model to be trained on fewer datapoints, as a single datapoint can be "reused" with a different style at each occurrence of that datapoint in training.

We proceed in Section \ref{sec:ourmodel} to introduce further key components of StyleMapper, including (1) the image style/content encoders and decoder, (2) the use of both raw and transformed images at each iteration of training to further encourage consistent style/content encoding and decoding operations, (3) image and style/content code reconstruction terms, and (4) a novel \textit{cross-domain reconstruction triplet loss} term that is used to encourage further generalization ability of the encoders and decoders for style transfer. We note that this method of generating unlimited possible style images in training could be adapted to train many other applicable style transfer methods within the medical imaging domain.

In Section \ref{sec:experiments} we explore experiments run with StyleMapper on the tasks of transferring test data to new target styles both simulated (Section \ref{sec:exp:oneshot}) and real (Section \ref{sec:exp:classifyscanners}), all while being trained on just 528 datapoints. We then discuss the limitations of this work in Section \ref{sec:limitations}, and conclude with Section \ref{sec:conclusion}.

The contribution of this paper can be summarized as follows.
\begin{enumerate}
    \item We introduce a new method for training arbitrary style transfer models with limited data within the medical imaging domain.
    \item We propose a new disentangled-representation learning style transfer model that uses this method, included a novel loss component.
    \item We demonstrate arbitrary style transfer and style discrimination on breast MRIs with our method, with both real and simulated medical imaging styles.
\end{enumerate}

\subsection*{Related Works}
\subsubsection*{Style Transfer in Natural Images}
Style transfer research in deep learning has often focused on transferring natural images to artistic styles. The seminal work of \cite{gatys2016image} surmounted the goal of arbitrary style transfer by leveraging the feature-extracting behavior that is intrinsic to convolutional neural networks \cite{krizhevsky2012imagenet}. In that work, the transfer between styles for a test image was performed by aligning the style information of the image with the information of the target style image. Later works expanded upon this idea with improvements such as greatly increased transfer speed \cite{adain}, the accounting for cross-channel correlations within image feature maps \cite{wct}, a closed form solution to the task \cite{closedform}, improved and more diverse stylization \cite{avatarnet}, and generally more robust stylization and content preservation \cite{multiadaptation, attentionmultistroke}.

However, these models are trained on tens of thousands of content (and in some cases, style) examples, via content and style datasets such as MS-COCO \cite{mscoco} and WikiArt \cite{wikiart}, respectively. This is not a problem within the aforementioned models' original context of \textit{artistic} style transfer, but if we wish to switch to the medical domain, obtaining similar quantities of usable, standardized training data can be very difficult in practice.
As such, we wish to develop an arbitrary style transfer method to be used within the medical imaging domain that can be trained on the lower end of typical sizes of many medical imaging datasets: only a few hundred images.

\subsubsection*{Style Transfer in Medical Images}
While a large portion of style transfer research focuses on artistic style transfer, there is still a rich literature of style transfer methods specialized for the domain of medical imaging. Following the many-to-many mapping approach of \cite{huang2018multimodal}, works such as \cite{yang2019unsupervised, yang2020cross} explore the adaptation of unpaired images across medical modality domains (CT and MR scans), also trained to utilize a shared content space with invertible mappings between image, style and content spaces. However, a key limitation of such methods is that they require observing the target test style in training and/or explicitly modifying the model architecture whenever an additional style is desired to be learned from and generalized to, something that StyleMapper does not require.

Other models have been built to automatically standardize different MRI image types (e.g. created by different manufacturers) \textit{without} explicitly providing knowledge of the underlying scanner technology that was used to generate the image, such as \cite{zhang2018automatic}, which used piecewise-linear mapping to normalize intensities across different anatomical regions. The study in \cite{yang2020mri} approaches the task of translating between different MRI modalities using Conditional GANs (\cite{mirza2014conditional}) and paired data. Unlike our approach and the other aforementioned disentangled representation-learning approaches, this method does not rely on the consistency of translating across image, content and style domains; instead it directly maps from one image space to another. Whereas our method forms separate estimates of style and content of test images, e.g. allowing for the interchange of styles for a fixed-content image, such style/content disentangling ability is not present in these works.

The work of \cite{modanwal2019normalization} uses CycleGANs (\cite{zhu2017cyclegan}) to learn normalization between breast MRIs of two different manufacturers, along with the addition of a mutual information loss term and a modified discriminator to ensure the consistency of intensity, noise and anatomical structure. As compared to traditional CycleGAN applications, the modifications of this method allow for training upon unpaired data, a philosophy that we follow due to the fact that cross-domain paired (medical) data is generally more scarce than unsorted data. However, this method is different from ours in that it cannot transfer to new styles unseen in training.

Similar to our goal of transferring images to a single fixed style is the work of \cite{ma2019neural}, which focuses on style transfer within the domain of 3D cardiovascular MRIs. Style transfer is performed in this work using hierarchical clustering methods to best map test images to the domain of training images, given the results of extracting features from various inner layers of a VGG-16 network. In this work, a test image is mapped to the most similar image out of the utilized training set using the Wasserstein distance, with style mapped according to a ``style library'' learned during training. A limitation of this is that rather than learning a fixed set/library of styles from the training set, we attempt to generalize our style encoder to be more flexible, and work on images of \textit{unseen} target styles that may only be vaguely similar to those seen in training.

\section{Materials and Methods}
\subsection{Dataset}
\label{sec:data}
For this work, we experimented with 628 breast MR (Magnetic Resonance) images taken from 628 different breast cancer patients with a GE Healthcare MRI machine, obtained from the Breast Cancer DCE-MRI dataset of \cite{ourdataset}.
All images have a $512\times512$ resolution, and were pre-processed by assigning the top 1\% of pixel values in the entire dataset to a value of 255, followed by linearly scaling the remaining pixel intensities to the 0-255 range, giving the data the same ``raw'' style. 528 datapoints were used to produce the training set, 50 were kept as a test set, and the other 50 were used for validation. 25 similarly preprocessed images from a Siemens scanner were also used in Section \ref{sec:exp:classifyscanners}. We describe the details of creating our specific dataset in Appendix \ref{app:data}.

\subsection{Methods}
In this section we will introduce a modified domain adaptation model \cite{yang2019unsupervised} and its evolution to our proposed model, StyleMapper.

\subsubsection{The Modified Baseline Model}
\label{sec:basemodel}
We begin with an unsupervised Domain Adaptation, Disentangled Representation-learning (DADR) model that can map between two different image domains by disentangling style and content representations within both of the domains \cite{huang2018multimodal, yang2019unsupervised}. In particular, given images $X_1$ and $X_2$ from different domains $\mathcal{X}_1$ and $\mathcal{X}_2$, respectively, the model can learn the representation of an image $X_i$ within a style space $\mathcal{S}_i$ and a content space $\mathcal{C}_i$ ($i=1,2$), described respectively via a style code $s_i$ and a content code $c_i$. We label this model as the \textit{baseline model}.

\paragraph{Diverse Styles via Image Transformation Functions}
\label{sec:transforms}
The baseline model performs well with style transfer to a set of discrete styles seen in training, but we wish to extend the work to transfer from an image of some style to an \textit{arbitrary} target style that is unseen in training. This is beneficial in the medical imaging field because many styles exist that may be desired to be transferred to, some of which have limited available data. As such, we propose to train the model on both raw images and style-transferred versions of the same images, with these styles simulated via random \textit{image transformation functions} that act on the raw images. In this way, the model learns to both distinguish between and extract different styles while keeping content unchanged. This not only allows for the model to adapt to a wider range of possible styles, but also allows the model to learn from fewer datapoints, because a single datapoint can be seen with a variety of distinct styles at different training iterations.

\begin{figure}[htbp]
\centering     
\subfigure[Original/linear]{\label{fig:a}\includegraphics[width=1in]{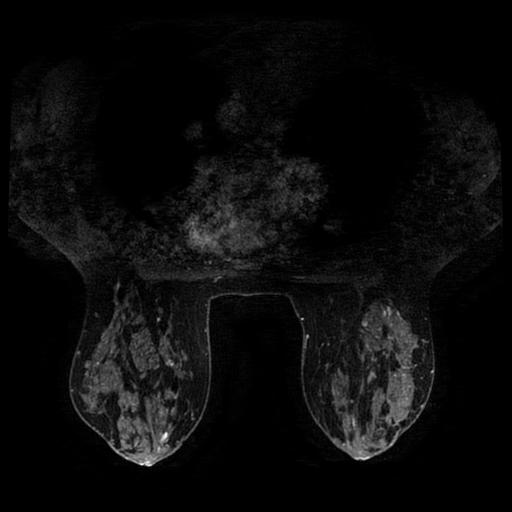}}
\subfigure[Negative]{\label{fig:b}\includegraphics[width=1in]{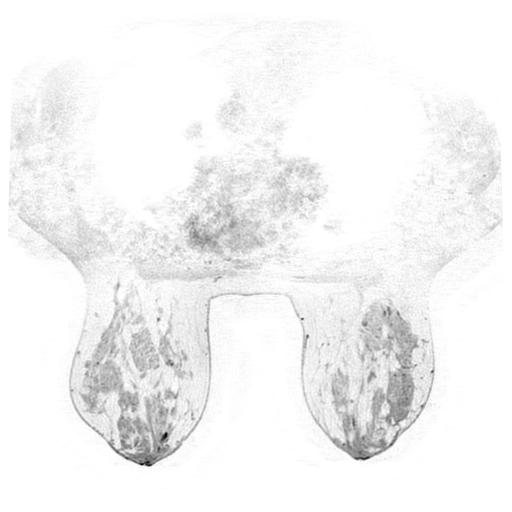}}
\subfigure[Log]{\label{fig:c}\includegraphics[width=1in]{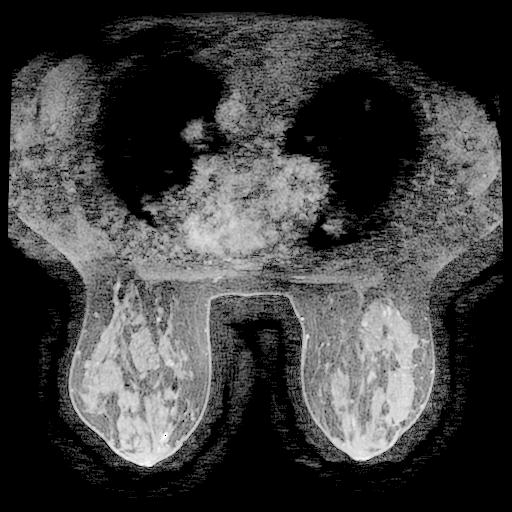}}
\subfigure[Power-Law (Gamma)]{\label{fig:d}\includegraphics[width=1in]{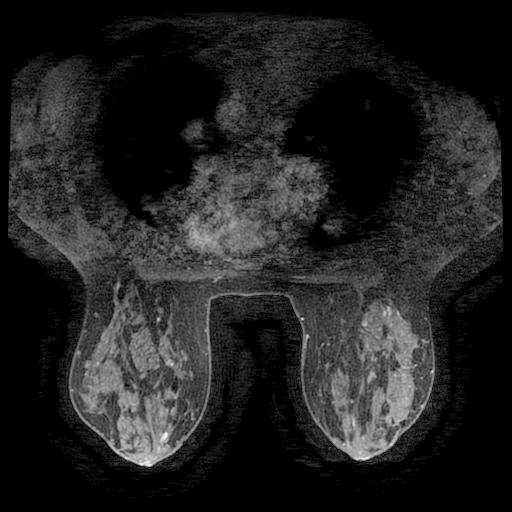}}
\subfigure[Piecewise-Linear]{\label{fig:e}\includegraphics[width=1in]{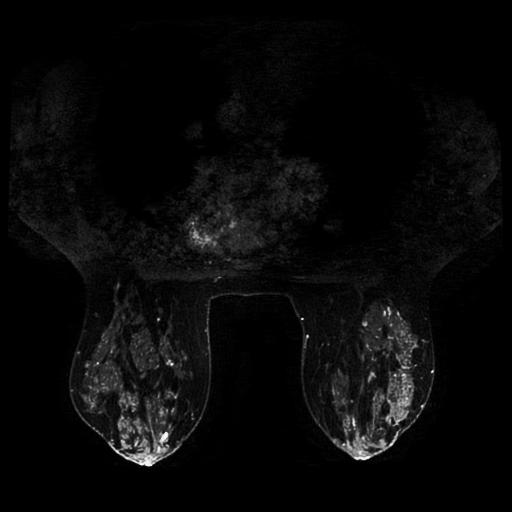}}
\subfigure[Sobel X]{\label{fig:f}\includegraphics[width=1in]{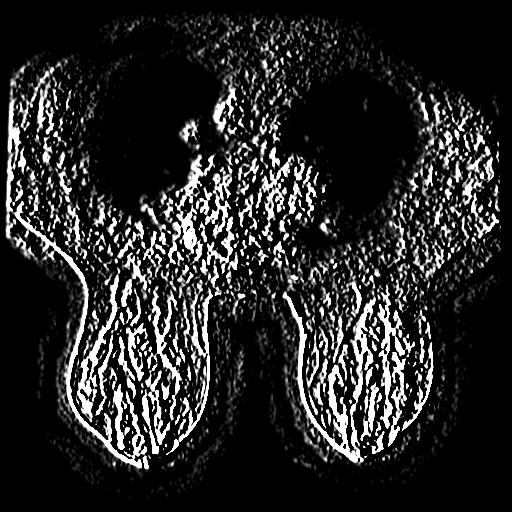}}
\subfigure[Sobel Y]{\label{fig:g}\includegraphics[width=1in]{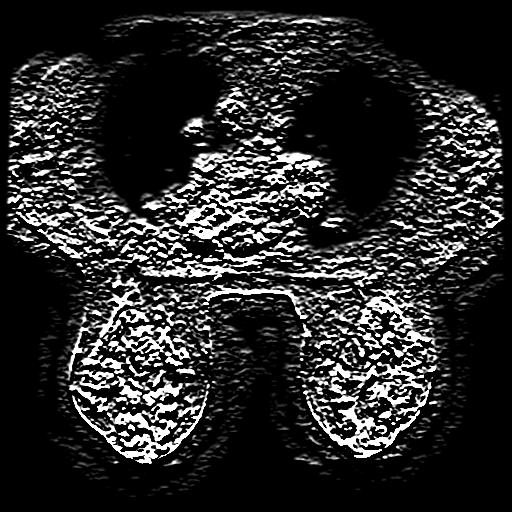}}
\caption{A comparison of the effects of the seven different image transformation functions that we use (Section \ref{sec:transforms} on a DCE-MRI breast scan, with randomized transfer function parameters fixed to the means of their sampling distributions.}
\label{fig:trans}
\end{figure}

In order to generate diverse styles for our model to learn, we use seven classes of some of the most common image intensity transformations \cite{transforms}: (1) the linear transformation, (2) the negative transformation, (3) the log transformation, (4) the power-law (gamma) transformation, (5) the piecewise-linear transformation, and (6,7) the Sobel X and Y operators. At each training step, one of the seven transformations are randomly selected to change a raw image to a new style. Although not very representative of the many possible \textit{artistic} styles of traditional style transfer works, we believe that the simulated styles described by the application of these random transformations, which manifest as generally nonlinear changes in pixel intensities, are a good proxy for many possible styles seen in medical imaging, which do not vary nearly as drastically as artistic styles do. We provide example images of each class of transformation in Figure \ref{fig:trans}.

The first five of these transformations are \textit{parametrically randomized}: when selected, the transformation function randomly selects its parameters from some pre-determined distribution. This allows the exact transfer function to be previously unseen at each training iteration.
The two-step randomized transformation function selection allows the model to extract codes corresponding to a practically unlimited range of distinct styles during training and to boost the style encoder's generalization ability and robustness at test time.

We provide the explicit formulae and probabilistic schema for generating the parameters of these transformation functions, as well as visual examples of them, in Appendix \ref{app:transforms}. We also take this simulated style approach for our experiments so that a ``ground truth'' deterministic transferred image can be directly compared to the neural style transfer result.

It is important to consider that the first five transformation functions are all \textit{invertible}, meaning that given an output pixel value $I_\mathrm{out}$, we can deterministically map $I_\mathrm{out}$ back to its corresponding input pixel value $I_\mathrm{in}$, implying that no information is lost through these transformations. The Sobel X and Y operators, however, could introduce information loss because of the additive nature of the convolution operation. In practice, because of the high resolution of the images, we assume that such an operation will only slightly affect the overall global content of the images, and thus decide to include the Sobel operators in the pool of possible transformations. Experiments of training without the Sobel operators further confirmed this statement. We do not ``randomize'' the Sobel transformations during training due to the more severe nature of these transformations as compared to the first five.

\paragraph{Modified Baseline Model Architecture}
\label{sec:baseline_architecture}
A precursor of our final model, the \textit{modified} baseline model consists of two main components: (1) content and style encoders for obtaining content and style representations, or \textit{codes}, of images and (2) generators/decoders for mapping content-style code pairs back to the space of images. In this model two encoder$\rightarrow$decoder$\rightarrow$encoder pipelines run in parallel: one for a raw image, and the other for the transformed version of the same image, where the image transformation is randomly chosen from one of the seven transformations described in Section \ref{sec:transforms}. 

The model is trained with objectives that enforce reconstruction where applicable both in-domain and cross-domain, as well as adversarial objectives that ensure the translated images to be indistinguishable from real images in the target domain.

\subsubsection{Central Model: StyleMapper}
\label{sec:ourmodel}
\begin{figure}[htbp]
\centering
\includegraphics[width=0.9\linewidth]{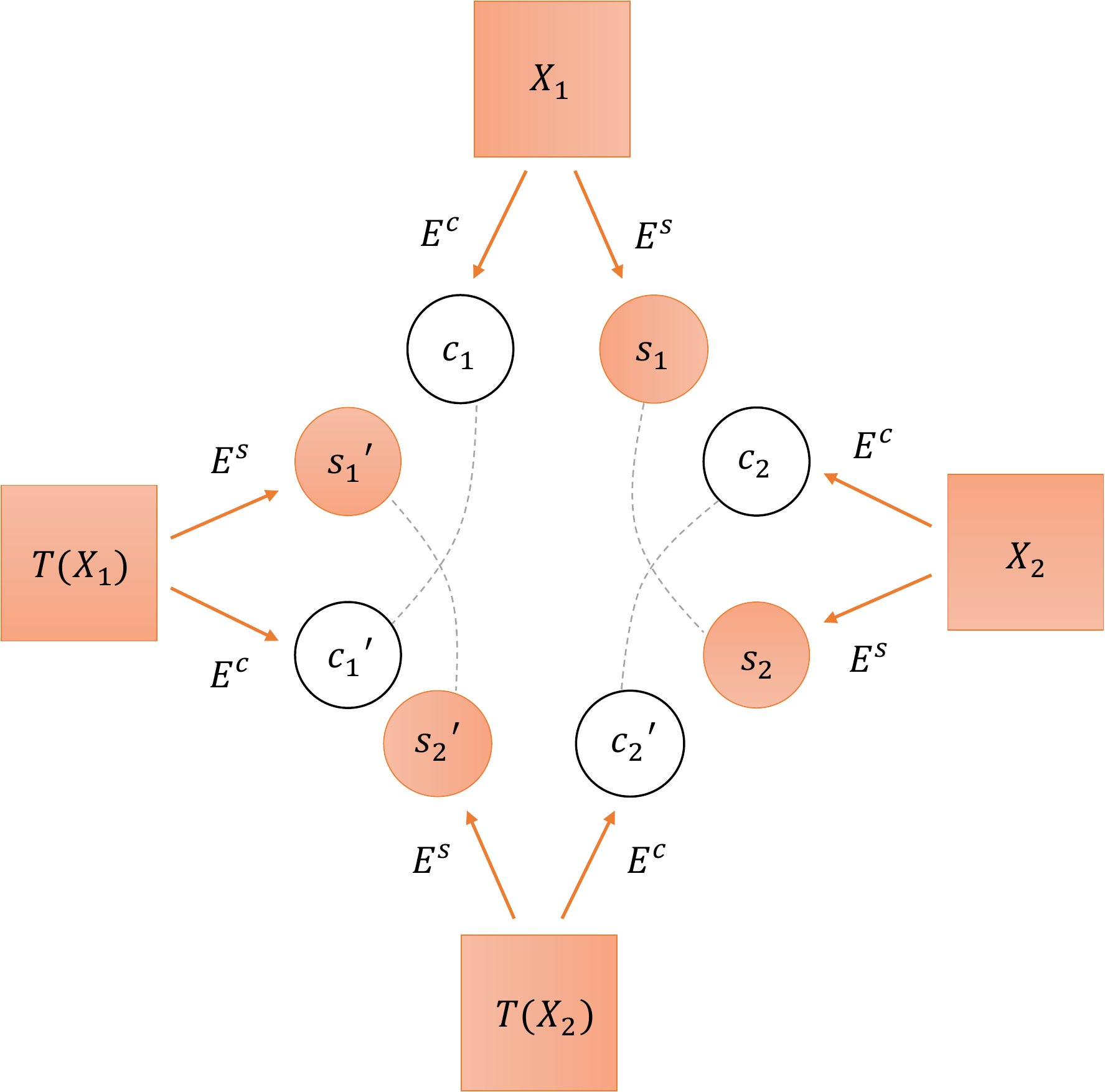}
\caption{\textbf{StyleMapper}: Our novel architecture used for style transfer. Solid arrows indicate encoding operations, and dashed lines indicate pairs of codes that should be optimally equivalent (Equation \eqref{eq:modelBsamelatentloss}), with the model trained to achieve as such. The decoder/generator $G$ is not pictured, as it receives input of various combinations of all of the pictured style and content codes.}
\label{fig:modelB}
\end{figure}

Using the modified baseline model of the previous section as a starting point, we created a custom style transfer model which we name StyleMapper (Figure \ref{fig:modelB}). We will next outline (1) the novel components of StyleMapper and (2) the main differences between StyleMapper and the modified baseline model, both in the architecture and in the training procedure/loss function. 
\paragraph{General Features}
\textbf{Multiple Data Pairs per Training Step.}
As opposed to the modified baseline model, StyleMapper is uses \textit{two} raw images per training iteration: a pair of distinct raw data 
$(X_1, X_2)$, and the results of applying the same random image transformation $T$ to that pair, $(T(X_1), T(X_2))$; $T$ is resampled for each pair $(X_1, X_2)$ at each training step. This implies that $X_1$ and $X_2$ should have the same style but different content, $X_i$ and $T(X_i)$ should have the same content but different style for $i=1,2$, and $T(X_1)$ and $T(X_2)$ should have the same style yet different content. As will be shown shortly, consistency of both content and style encoding can then be enforced through reconstruction constraints of different pairings of these four images that should have the same content or style, respectively, further encouraging the encoders and decoder to work across diverse domains.
\textbf{Unifying the Encoders and Decoders.}
In the modified baseline model, separate content and style encoder/decoder groups are trained corresponding to each of the input images and their corresponding transformed versions. We must consider the possibility of there being inconsistencies between the members of each of these pairs of networks; to account for this, we switch to using single encoders for content and for style $E^c, E^s$ and a single decoder/generator $G$, allowing for the potential of a boost in style transfer generalizability, and a simpler model.

\textbf{Most Representative Style Code.}
Upon inference, we introduce a fixed most representative style code $s$ instead of a target style code (as in \cite{yang2019unsupervised}) to map input images to the target style $\overline{s}$ is defined as the style code that is closest on average to all of the style codes of that test set. Specifically, for each of the $N$ images $X_{i}$ in the target style test set, record the style code $s_i$ obtained from the trained model's style encoder. If $N>1$, we then compute the most representative style code as
\begin{equation}
\label{eq:mostrep}
     \overline{s} = s_{k} \quad\mathrm{where}\quad
     k = \underset{i}{\mathrm{argmin}}\frac{1}{N-1}\sum_{j: j\neq i}^{N-1}\mathrm{MAE}(s_i, s_j),
\end{equation}
and $\mathrm{MAE}(\cdot, \cdot)$ is the mean absolute error function. In the case of $N=1$, $\overline{s}$ is simply the style code obtained from the single target style image. This task is labeled as few-shot style transfer. In Section \ref{sec:exp:oneshot}, we show that $N=1$ is sufficient for successful style transfer, meaning that our method is compatible to one-shot learning.

\paragraph{Training Loss Terms}
\textbf{Image Reconstruction Loss.}
The model should be able to reconstruct an image sampled from the data distribution after encoding and decoding. To achieve this, the style and content encoders $E^s, E^c$ and decoder $G$ are trained to minimize the mean absolute error/$L_1$ distance between reconstructed images and original images, via the image reconstruction loss from \cite{yang2019unsupervised} of 
\begin{equation}
\label{eq:loss_imgrecon}
\mathcal{L}_\mathrm{recon}(X_1) = \mathop{\mathbb{E}}\limits_{X_1\sim p(X_1)}\left[\left\Vert G\left(E^c(X_1), E^s(X_1)\right)-X_1)\right\Vert_1\right],
\end{equation}
where $p(X_1)$ is the distribution of $X_1$ data. Further image reconstruction loss terms $\mathcal{L}_\mathrm{recon}(X_2)$, $\mathcal{L}_\mathrm{recon}(T(X_1))$ and $\mathcal{L}_\mathrm{recon}(T(X_2))$ for $X_2$, $T(X_1)$ and $T(X_2)$ are then respectively defined the same. 

We note that when building StyleMapper from the baseline model, we removed the discriminator because the generator can solely be trained by the image reconstruction loss, and we do not need the discriminator for any classification role. As such, the adversarial loss term of the modified baseline model (\cite{yang2019unsupervised}) is not present for StyleMapper.

\textbf{Latent Reconstruction Loss.}
We wish to encourage translation and reconstruction across diverse domains of style and content. One characteristic of StyleMapper that is different from the modified baseline model is that the (encode$\rightarrow$decode$\rightarrow$encode) progression found in the modified baseline model used for latent space reconstruction is reduced to (encode$\rightarrow$decode), such that we no longer train for latent space consistency in this manner. 

Instead, we enforce these latent embedding consistency requirements with style and content reconstruction loss terms, adapted from \cite{yang2019unsupervised}, that are defined respectively as
\begin{equation}
\begin{split}
\label{eq:modelBsamelatentloss}
\mathcal{L}_{\mathrm{same}_{s}} &=  \mathbb{E}\left[\left\Vert E^s(X_1)-E^s(X_2)\right\Vert_1\right] \\
\mathcal{L}_{\mathrm{same}_{s:T}} &= \mathbb{E}\left[\left\Vert E^s(T(X_1))-E^s(T(X_2))\right\Vert_1\right] \\
\mathcal{L}_{\mathrm{same}_{c: X_i}} &= \mathbb{E}\left[\left\Vert E^c(X_i)-E^c(T(X_i))\right\Vert_1\right],
\end{split}
\end{equation}
with $i=1,2$. These constraints can be seen via the dashed lines in Figure \ref{fig:modelB}.

\textbf{Cross-Domain Reconstruction Triplet Loss.}
We include a \textit{cross-domain reconstruction triplet loss} term, to encourage content and/or style reconstruction given twelve certain combinations of the images $X_1, X_2, T(X_1)$ and $T(X_2)$, as
\begin{equation}
\label{eq:loss_cross}
\mathcal{L}_{\mathrm{cross}} = \sum_{\displaystyle(p_1, p_2, p_3)\in\mathcal{P}}\mathbb{E}\left[\left\Vert G\left(E^c(p_1), E^s(p_2)\right)-p_3\right\Vert_1\right],
\end{equation}
where $\mathcal{P}$ is the set of twelve triplets constructed from $p_1, p_2, p_3\in \{X_1, X_2, T(X_1), T(X_2)\}$ by the condition
\begin{equation}
    \mathcal{P} = \left\{(p_1,p_2,p_3):E^c(p_1)=E^c(p_3), E^s(p_2)=E^s(p_3)  \right\}
\end{equation}
(note that $p_1, p_2$ and $p_3$ don't necessarily have to be different images). This loss term is important for training the encoders and decoder to have flexible and generalizable performance across domains, and is written explicitly in Appendix \ref{app:fullcrossloss}.

We now come to the full loss function that is minimized to train StyleMapper,
\begin{equation}
\label{eq:lossB}
\begin{split}
\mathcal{L}_{\mathrm{StyleMapper}} &= \lambda_{\mathrm{recon}}\left[ \mathcal{L}_\mathrm{recon}(X_1) + \mathcal{L}_\mathrm{recon}(X_2)\right.\\ &+\left.\mathcal{L}_\mathrm{recon}(T(X_1)) + \mathcal{L}_\mathrm{recon}(T(X_2))\right]\\
        & + \lambda_{\mathrm{same}_s}\left(\mathcal{L}_{\mathrm{same}_{s}} + \mathcal{L}_{\mathrm{same}_{s:T}}\right) \\
        & + \lambda_{\mathrm{same}_c}\left(\mathcal{L}_{\mathrm{same}_{c: X_1}} + \mathcal{L}_{\mathrm{same}_{c: X_2}}\right) \\
        &+ \lambda_{\mathrm{cross}}\mathcal{L}_{\mathrm{cross}},
\end{split}
\end{equation}
where $\lambda_{\mathrm{recon}}$, $\lambda_{\mathrm{cross}}$, $\lambda_{\mathrm{same}_s}$, and $\lambda_{\mathrm{same}_c}$ are loss weight hyperparameters. We will now proceed to implementational and training details in the next section, followed by our experimental results in Section \ref{sec:experiments}.

\subsubsection{Implementational Details}
\label{sec:implementation}
\textbf{Network Architecture.}
We build off of the MUNIT model of \cite{huang2018multimodal}. Content encoders consist of several strided convolutional layers to downsample the input, and several residual blocks to further process it. All convolutional layers are followed by Instance Normalization (IN) modules \cite{ulyanov2016instance}. Style encoders include several strided convolutional layers, followed by a global average pooling layer and a fully connected (FC) layer. We do not use IN layers in the style encoder, since IN removes the original feature mean and variance that represent important style information.

\textbf{Network Training.}
We use the Adam optimizer \cite{kingma2014adam} to train StyleMapper by minimizing the loss (Equation \eqref{eq:lossB}), with weight decay strength of 0.0001, $\beta_1=0.5$ and $\beta_2=0.999$. Kaiming's Method \cite{he2015delving} was used to initialize model weights, and we trained with a learning rate of 0.0001. We used a batch size of 1 (due to memory limitations), training until no further minimization of the loss term(s) was observed (a minimum of about 10,000 iterations was needed in essentially all cases). 

For the loss weights, we used values of $\lambda_\mathrm{recon}=10$, $\lambda_{\mathrm{same}_c}=\lambda_{\mathrm{same}_s}=5$ and $\lambda_{\mathrm{cross}}=1$. Additional hyperparameters from MUNIT are unchanged from their settings in that work. We train our models until we observe loss convergence, assisted by validation via the MAE residuals between style transferred image results through learning style mapping (our model) and transferred image results through direct image transformations (the ``ground truth'' to compare to). All computations are performed with an NVIDIA QUADRO M6000 24GB GPU.

\section{Results}
\label{sec:experiments}
\subsection{One-shot Style Transfer I: Simulated Styles}
\label{sec:exp:oneshot}
The core goal of our StyleMapper model is to be able to transfer a test image to some unseen target style while preserving content. To test this, we train StyleMapper following Equation \eqref{eq:lossB} on all image transformation functions/styles in Section \ref{sec:transforms}, but \textit{excluding} a particular class of transformation $T$ (with parameters fixed) from the pool of possible transformations, to be used as a target style.
After training, we apply $T$ to the first 25 of the test set to obtain $\{T(X_{\mathrm{target}})\}$, and then extract the content and style codes of each of these $T(X_{\mathrm{target}})$ using the style encoder to obtain codes $\{s^T_{\mathrm{target}}\}$. Finally, we obtain a most representative style code $\overline{s}^T$ for the target style by applying Equation \eqref{eq:mostrep} to some $N_\mathrm{target}$ of these 25 style codes, to judge how many target style images the model needs to observe to compute a useful target style code.

We evaluate the ability of the style encoder to extract the correct style code from the target style test images by taking the remaining 25 of the test set $\{X_{\mathrm{test}}\}$, transferring these images to the target style via $\overline{s}^T$ to obtain $\{X^{s:T}_{\mathrm{test}}\}$, and comparing these to the ``ground truth'' of transformed images $\{T(X_{\mathrm{test}})\}$. In particular, the content encoder extracts content codes $\{c_{\mathrm{test}}\}$ from the images $\{X_{\mathrm{test}}\}$, and the generator/decoder $G$ takes each of these content codes with the target style code $\overline{s}^T$ to synthesize the transferred images $\left\{G(c_{\mathrm{test}}, \overline{s}^T) : c_{\mathrm{test}} \in \{c_{\mathrm{test}}\} \right\} = \{X^{s:T}_{\mathrm{test}}\}$.

We do this comparison using the MAE between $\{X^{s:T}_{\mathrm{test}}\}$ and $\{T(X_{\mathrm{test}})\}$. We also note that for better performance comparison between styles, for a given style we normalize all MAEs across the different $N_\mathrm{target}$ values by dividing each MAE by $\mathrm{MAE}\left(\{X_{\mathrm{test}}\}, \{T(X_{\mathrm{test}})\}\right)$. 

We will explore examples of this by transferring to (1) a target style that is fairly similar to those seen in training--the log transformation for a model trained on all transformations \textit{but} log, and the same but for the gamma/power-law transformation--and (2) a target style that is distinct from the styles/transformation seen in training (the exponential function $\exp(\cdot)$). We test these on a range of $N_\mathrm{target}$ to see if the most representative target style computation is dependent on the quantity of target style data seen by the style encoder.

To begin we examine a log target style, which we define via the logarithmic intensity transfer function with parameter fixed to its average value (Equation \eqref{eq:trans_log} in Appendix \ref{app:transforms}). 
Next we perform the same experiments with a power-law target style, via the power-law transfer function with exponential parameter fixed to $\tilde{\gamma}=0.5$ (Equation \eqref{eq:trans_gamma} in Appendix \ref{app:transforms}). Finally, we test a target style described by the exponential transfer function equation $T(I_\mathrm{in})=a\exp(bI_\mathrm{in})$ with $a=2.3, b=0.02$, on a style encoder trained on \textit{all} of the parametrically-random styles of Section \ref{sec:transforms}. In this case, we have a style with a transfer function curve that is not as similar to any of the possible randomized transfer curves seen by the style encoder during training as in the first two examples, where the target log and power-law transformations had the possibility of being similar to certain settings of the randomized power-law and log transformations seen in training, respectively.

\begin{figure*}
\centering
\includegraphics[width=0.95\linewidth]{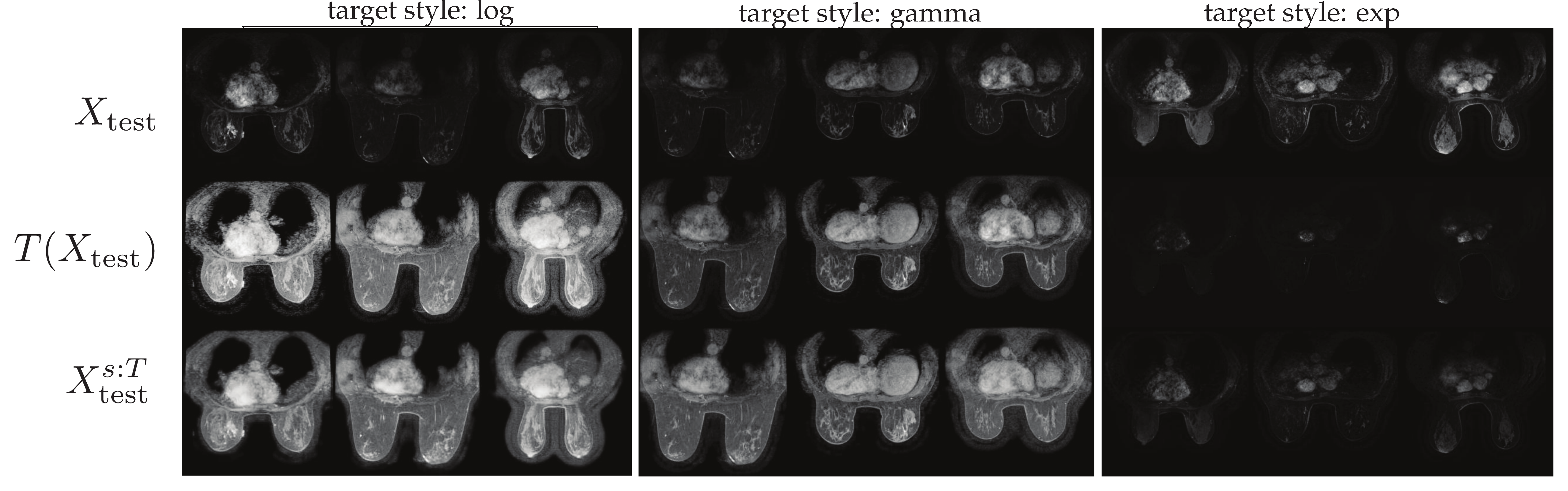}
\caption{\textbf{One-shot style transfer with various target styles: Qualitative Results.} See Section \ref{sec:exp:oneshot}. Transferring a set of 25 MR test images $\{X_{\mathrm{test}}\}$ (top row) to different target styles not seen in training $\{X^{s:T}_{\mathrm{test}}\}$ (bottom row), with target style code obtained from a \textbf{single test image of the target style}. The transferred images are compared to the ``ground truth'' $\{T(X_{\mathrm{test}})\}$ (middle row) of the images directly transformed by the target style's corresponding transformation function $T$. From left to right, the target styles/transformations are the fixed log, gamma/power-law and exp transformations, as described in Section \ref{sec:exp:oneshot}, and for each style, three random images are visualized. Accompanying quantitative results in Figure \ref{fig:fewshot_quant}.}
\label{fig:fewshot_qual}
\end{figure*}

\begin{figure*}
\centering
\includegraphics[width=0.95\linewidth]{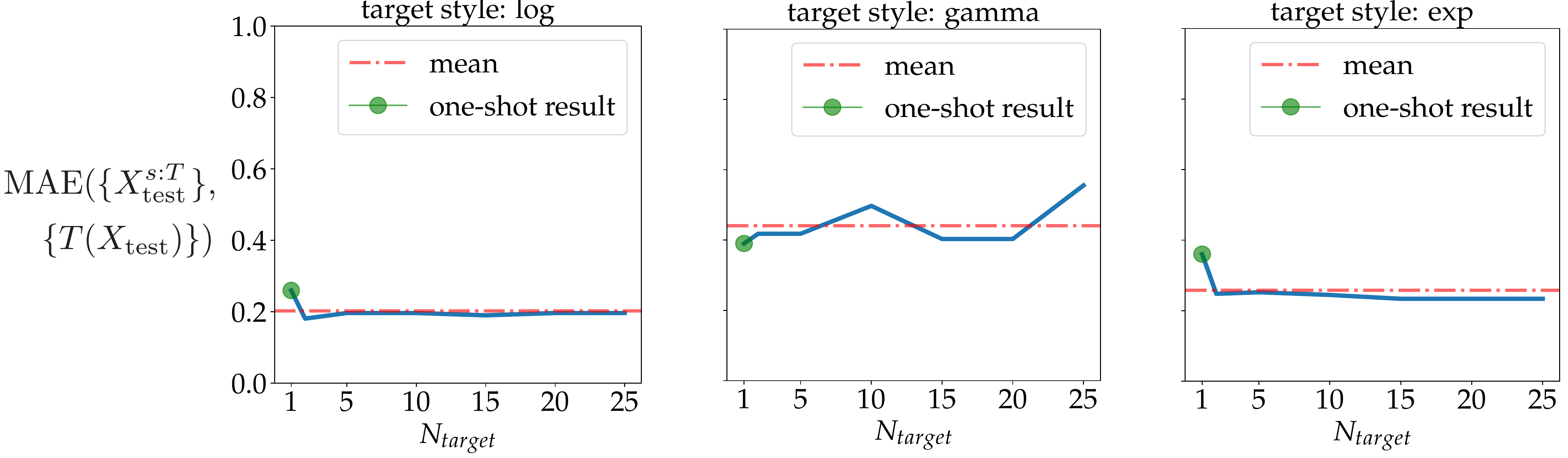}
\caption{\textbf{One-shot style transfer with various target styles: Quantitative Results.} See Section \ref{sec:exp:oneshot}. Mean absolute error (MAE) between style transferred images $\{X^{s:T}_{\mathrm{test}}\}$ and ``ground truth'' transformed images $\{T(X_{\mathrm{test}})\}$, indicating performance of style transfer, with respect to number of target style images $N_\mathrm{target}$ used to compute the most representative target style code that is used to perform style transfer. Accompanying qualitative results in Figure \ref{fig:fewshot_qual}.}
\label{fig:fewshot_quant}
\end{figure*}

The qualitative and quantitative results of these experiments are shown in Figures \ref{fig:fewshot_qual} and \ref{fig:fewshot_quant}, respectively. We see that style transfer performance described by the MAE between $\{X^{s:T}_{\mathrm{test}}\}$ and $\{T(X_{\mathrm{test}})\}$ is mostly independent to $N_\mathrm{target}$, implying that only a single target style image is needed by the style encoder to perform style transfer. In particular, for this one-shot case, we have MAEs for the log, gamma and exp styles of 0.2595, 0.3902 and 0.3601, respectively.  As explored in Appendix \ref{app:stylecodes}, indeed the style codes extracted from different images of one same style are almost identical, implying that the most-representative style code obtained from aggregating $N_\mathrm{target}$ of these individual codes will be almost the same as any one of them.

The behavior of the style encoder over a range of styles, as well as the structure of the style codes themselves, are worth exploring. Although beyond the scope of the main body of this work, we explore both how extracted style codes differ between (1) different styles and (2) different images of the same style in Appendix \ref{app:stylecodes}.

\subsection{One-shot Style Transfer II: MRI Scanner Styles}
\label{sec:exp:classifyscanners}
We will now explore the ability of StyleMapper to transfer images to a new medical scanner style that is unseen in training, in particular the real style Siemens MR scanners, as GE scanner data was used to train the model, while Siemens data has never been observed.

\begin{figure}
\centering
\includegraphics[width=0.9\linewidth]{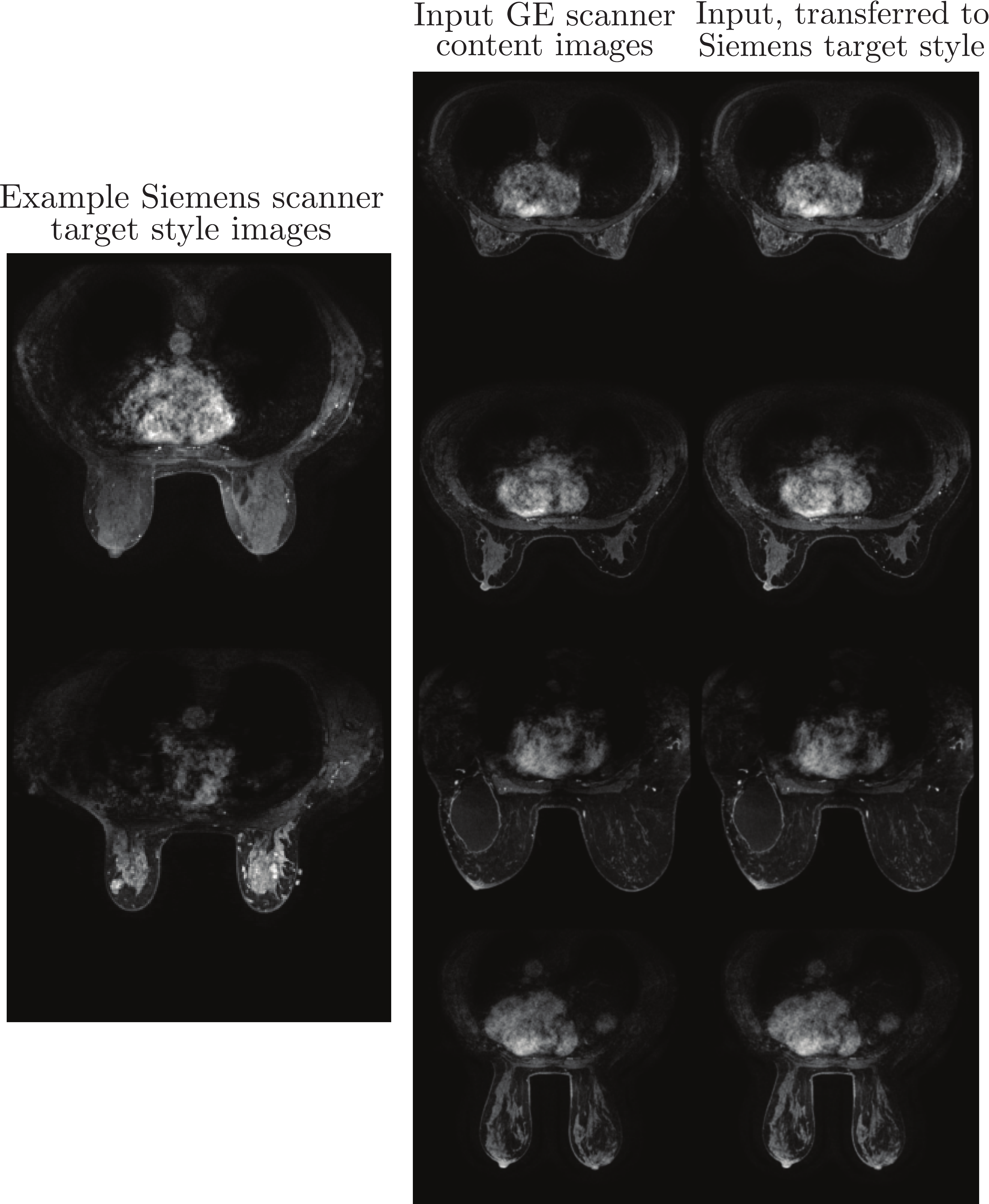}
\caption{\textbf{MRI Style Transfer to Unseen Scanner Style.} Results (right column) of transferring GE scanner MR Images (center column) to the Siemens scanner style unseen in training (left column).}
\label{fig:GE_to_Siemens}
\end{figure}

We performed one-shot style transfer on the same set of 25 raw GE scanner images as in the previous section, with a StyleMapper trained on all of the randomized styles (Section \ref{sec:transforms}), and a single Siemens scan image used to obtain the target style. Example results of this are shown in Figure \ref{fig:GE_to_Siemens}. Given that there is no ``ground truth'' to compare the transferred results to as in the previous section of simulated target styles, we believe these results to be strong given the fact that certain stylistic characteristics of Siemens scans as compared to GE scans--such as Siemens on average appearing to be slightly brighter than GE--appear in the transferred results.

\begin{figure}
\centering
\includegraphics[width=1.0\linewidth]{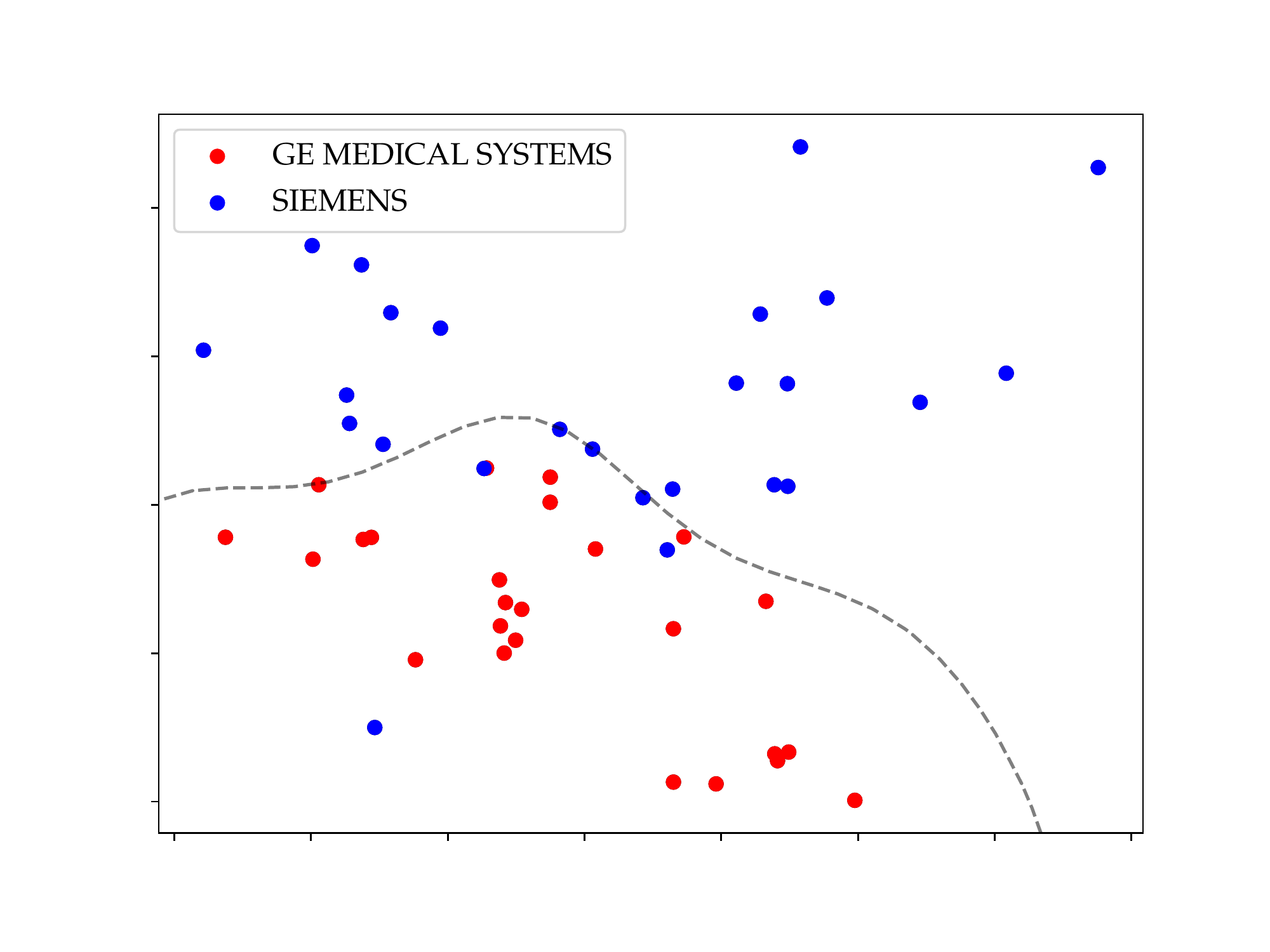}
\caption{\textbf{Discriminating a realistic unseen style.} Using a StyleMapper style encoder we extracted style codes of a set of unpaired MR scans of two different manufacturers, GE and Siemens, with the latter style previously unseen by the model. Pictured are these style codes embedded into $\mathbb{R}^2$, and the decision boundary learned by training a support vector classifier (SVC) on them. Classification accuracy: 88.0\%. Figure recommended to be viewed in color.}
\label{fig:classifyscanners}
\end{figure}

We can also use the same StyleMapper to distinguish between these two styles of GE and Siemens. To do this, we use the style encoder of StyleMapper to extract style codes from 25 GE and 25 Siemens images (all with different content). Next, we performed dimensionality reduction on these 8-dimensional style code vectors via principal component analysis (PCA) to map them to $\mathbb{R}^2$, and trained a support vector classifier (SVC) with a radial basis function kernel to discriminate between the two classes of points \cite{svc}.
As shown in Figure \ref{fig:classifyscanners}, the style encoder is able to usually discriminate successfully between the two styles via the differences in their encodings, with an SVC accuracy of $88.0\%$.

\subsection{Ablation Study: Finite Set of Training Styles}
\label{sec:exp:ablation}

In order to show the necessity of using parametrically-randomized transformations as training styles, rather than the fixed transformations tested in Appendix \ref{app:stylecodes}--we will attempt one of the same few-shot style transfer experiments of Section \ref{sec:exp:oneshot}, but with a style encoder trained only on these fixed transformations. In other words, the former configuration allows for the style encoder to see a continuous range of styles in training--technically a new particular style at every iteration (excluding the Sobel transformations)--while the latter only gives the style encoder a discrete, finite set of possible styles to learn to extract style codes from, a problem that is exacerbated when only limited training data is present.

We will repeat the experiment with the same fixed-parameter log target style as in Section \ref{sec:exp:oneshot}, but with a model trained on \textit{fixed versions} of all of the other six transformations, with parameters fixed to their average values (Appendix \ref{app:stylecodes}), except for the power-law function which is fixed the same as in the target power-law style experiment of Section \ref{sec:exp:oneshot}. The failure of using this finite set of transformations is seen when comparing the one-shot transfer and MAE between the transformed ``ground truth'' and the transferred result, of 0.2778, to be compared to 0.1178 for the non-ablated case (Section \ref{sec:exp:oneshot}).

Just as in the compared experiment, we found the MAE here to not be improved by increasing $N_\mathrm{target}$, the number of target style images used by the style encoder to estimate the target style code used to perform transfer. We note that we observed significantly more noise on MAE values about the one-shot transfer MAE with respect to $N_\mathrm{target}$ than in the other experiment, indicating that the style encoder was not nearly as consistent as for the case of it being trained on parametrically-random transformations, extracting erroneous, but different codes from the target style images.

\section*{Discussion}
\label{sec:limitations}
One limitation of our work is that in order to successfully estimate the correct target style code, test target styles usually need to be at least somewhat similar to the styles seen in training; not identical, but also not completely orthogonal. Target styles that are very distant from those seen in training, that require the model to perform large amounts of \textit{extrapolation}, not just \textit{interpolation}, give more trouble.

We focused on training the model on simulated styles described by intensity transfer functions, in order to focus on content-preserving medical imaging styles and to facilitate training on a small dataset. However, an important future endeavor will be to explore how to train and test the model on non-medical images of more diverse styles, to see how well it can generalize to these situations, while potentially maintaining the requirement for only limited data.
\section*{Conclusions}
\label{sec:conclusion}
In this work we introduced a novel medical image style transfer method, StyleMapper, that can transfer images to a new target style unseen in training while observing only a single image of this style at test time, and can be successfully trained on limited amounts of single-style data. We explored the applications of StyleMapper to both style transfer and the classification of unseen styles.

\backmatter

\bmhead{Supplementary information}

This article has accompanying supplementary appendices, that describe details of the dataset, the image transformation functions/training styles, additional details for the novel loss function, and further experiments.

%

\begin{appendices}






\end{appendices}


\bibliography{styletransfer}


\begin{thebibliography}{25}
\ifx \bisbn   \undefined \def \bisbn  #1{ISBN #1}\fi
\ifx \binits  \undefined \def \binits#1{#1}\fi
\ifx \bauthor  \undefined \def \bauthor#1{#1}\fi
\ifx \batitle  \undefined \def \batitle#1{#1}\fi
\ifx \bjtitle  \undefined \def \bjtitle#1{#1}\fi
\ifx \bvolume  \undefined \def \bvolume#1{\textbf{#1}}\fi
\ifx \byear  \undefined \def \byear#1{#1}\fi
\ifx \bissue  \undefined \def \bissue#1{#1}\fi
\ifx \bfpage  \undefined \def \bfpage#1{#1}\fi
\ifx \blpage  \undefined \def \blpage #1{#1}\fi
\ifx \burl  \undefined \def \burl#1{\textsf{#1}}\fi
\ifx \doiurl  \undefined \def \doiurl#1{\url{https://doi.org/#1}}\fi
\ifx \betal  \undefined \def \betal{\textit{et al.}}\fi
\ifx \binstitute  \undefined \def \binstitute#1{#1}\fi
\ifx \binstitutionaled  \undefined \def \binstitutionaled#1{#1}\fi
\ifx \bctitle  \undefined \def \bctitle#1{#1}\fi
\ifx \beditor  \undefined \def \beditor#1{#1}\fi
\ifx \bpublisher  \undefined \def \bpublisher#1{#1}\fi
\ifx \bbtitle  \undefined \def \bbtitle#1{#1}\fi
\ifx \bedition  \undefined \def \bedition#1{#1}\fi
\ifx \bseriesno  \undefined \def \bseriesno#1{#1}\fi
\ifx \blocation  \undefined \def \blocation#1{#1}\fi
\ifx \bsertitle  \undefined \def \bsertitle#1{#1}\fi
\ifx \bsnm \undefined \def \bsnm#1{#1}\fi
\ifx \bsuffix \undefined \def \bsuffix#1{#1}\fi
\ifx \bparticle \undefined \def \bparticle#1{#1}\fi
\ifx \barticle \undefined \def \barticle#1{#1}\fi
\bibcommenthead
\ifx \bconfdate \undefined \def \bconfdate #1{#1}\fi
\ifx \botherref \undefined \def \botherref #1{#1}\fi
\ifx \url \undefined \def \url#1{\textsf{#1}}\fi
\ifx \bchapter \undefined \def \bchapter#1{#1}\fi
\ifx \bbook \undefined \def \bbook#1{#1}\fi
\ifx \bcomment \undefined \def \bcomment#1{#1}\fi
\ifx \oauthor \undefined \def \oauthor#1{#1}\fi
\ifx \citeauthoryear \undefined \def \citeauthoryear#1{#1}\fi
\ifx \endbibitem  \undefined \def \endbibitem {}\fi
\ifx \bconflocation  \undefined \def \bconflocation#1{#1}\fi
\ifx \arxivurl  \undefined \def \arxivurl#1{\textsf{#1}}\fi
\csname PreBibitemsHook\endcsname

\bibitem{ma2019neural}
\begin{bchapter}
\bauthor{\bsnm{Ma}, \binits{C.}},
\bauthor{\bsnm{Ji}, \binits{Z.}},
\bauthor{\bsnm{Gao}, \binits{M.}}:
\bctitle{Neural style transfer improves 3d cardiovascular mr image segmentation
  on inconsistent data}.
In: \bbtitle{International Conference on Medical Image Computing and
  Computer-Assisted Intervention},
pp. \bfpage{128}--\blpage{136}
(\byear{2019}).
\bcomment{Springer}
\end{bchapter}
\endbibitem

\bibitem{gatys2016image}
\begin{bchapter}
\bauthor{\bsnm{Gatys}, \binits{L.A.}},
\bauthor{\bsnm{Ecker}, \binits{A.S.}},
\bauthor{\bsnm{Bethge}, \binits{M.}}:
\bctitle{Image style transfer using convolutional neural networks}.
In: \bbtitle{Proceedings of the IEEE Conference on Computer Vision and Pattern
  Recognition},
pp. \bfpage{2414}--\blpage{2423}
(\byear{2016})
\end{bchapter}
\endbibitem

\bibitem{krizhevsky2012imagenet}
\begin{barticle}
\bauthor{\bsnm{Krizhevsky}, \binits{A.}},
\bauthor{\bsnm{Sutskever}, \binits{I.}},
\bauthor{\bsnm{Hinton}, \binits{G.E.}}:
\batitle{Imagenet classification with deep convolutional neural networks}.
\bjtitle{Advances in neural information processing systems}
\bvolume{25},
\bfpage{1097}--\blpage{1105}
(\byear{2012})
\end{barticle}
\endbibitem

\bibitem{adain}
\begin{bchapter}
\bauthor{\bsnm{Huang}, \binits{X.}},
\bauthor{\bsnm{Belongie}, \binits{S.}}:
\bctitle{Arbitrary style transfer in real-time with adaptive instance
  normalization}.
In: \bbtitle{Proceedings of the IEEE International Conference on Computer
  Vision},
pp. \bfpage{1501}--\blpage{1510}
(\byear{2017})
\end{bchapter}
\endbibitem

\bibitem{wct}
\begin{botherref}
\oauthor{\bsnm{Li}, \binits{Y.}},
\oauthor{\bsnm{Fang}, \binits{C.}},
\oauthor{\bsnm{Yang}, \binits{J.}},
\oauthor{\bsnm{Wang}, \binits{Z.}},
\oauthor{\bsnm{Lu}, \binits{X.}},
\oauthor{\bsnm{Yang}, \binits{M.-H.}}:
Universal style transfer via feature transforms.
arXiv preprint arXiv:1705.08086
(2017)
\end{botherref}
\endbibitem

\bibitem{closedform}
\begin{bchapter}
\bauthor{\bsnm{Lu}, \binits{M.}},
\bauthor{\bsnm{Zhao}, \binits{H.}},
\bauthor{\bsnm{Yao}, \binits{A.}},
\bauthor{\bsnm{Chen}, \binits{Y.}},
\bauthor{\bsnm{Xu}, \binits{F.}},
\bauthor{\bsnm{Zhang}, \binits{L.}}:
\bctitle{A closed-form solution to universal style transfer}.
In: \bbtitle{Proceedings of the IEEE/CVF International Conference on Computer
  Vision},
pp. \bfpage{5952}--\blpage{5961}
(\byear{2019})
\end{bchapter}
\endbibitem

\bibitem{avatarnet}
\begin{bchapter}
\bauthor{\bsnm{Sheng}, \binits{L.}},
\bauthor{\bsnm{Lin}, \binits{Z.}},
\bauthor{\bsnm{Shao}, \binits{J.}},
\bauthor{\bsnm{Wang}, \binits{X.}}:
\bctitle{Avatar-net: Multi-scale zero-shot style transfer by feature
  decoration}.
In: \bbtitle{Proceedings of the IEEE Conference on Computer Vision and Pattern
  Recognition},
pp. \bfpage{8242}--\blpage{8250}
(\byear{2018})
\end{bchapter}
\endbibitem

\bibitem{multiadaptation}
\begin{bchapter}
\bauthor{\bsnm{Deng}, \binits{Y.}},
\bauthor{\bsnm{Tang}, \binits{F.}},
\bauthor{\bsnm{Dong}, \binits{W.}},
\bauthor{\bsnm{Sun}, \binits{W.}},
\bauthor{\bsnm{Huang}, \binits{F.}},
\bauthor{\bsnm{Xu}, \binits{C.}}:
\bctitle{Arbitrary style transfer via multi-adaptation network}.
In: \bbtitle{Proceedings of the 28th ACM International Conference on
  Multimedia},
pp. \bfpage{2719}--\blpage{2727}
(\byear{2020})
\end{bchapter}
\endbibitem

\bibitem{attentionmultistroke}
\begin{bchapter}
\bauthor{\bsnm{Yao}, \binits{Y.}},
\bauthor{\bsnm{Ren}, \binits{J.}},
\bauthor{\bsnm{Xie}, \binits{X.}},
\bauthor{\bsnm{Liu}, \binits{W.}},
\bauthor{\bsnm{Liu}, \binits{Y.-J.}},
\bauthor{\bsnm{Wang}, \binits{J.}}:
\bctitle{Attention-aware multi-stroke style transfer}.
In: \bbtitle{Proceedings of the IEEE/CVF Conference on Computer Vision and
  Pattern Recognition},
pp. \bfpage{1467}--\blpage{1475}
(\byear{2019})
\end{bchapter}
\endbibitem

\bibitem{mscoco}
\begin{bchapter}
\bauthor{\bsnm{Lin}, \binits{T.-Y.}},
\bauthor{\bsnm{Maire}, \binits{M.}},
\bauthor{\bsnm{Belongie}, \binits{S.}},
\bauthor{\bsnm{Hays}, \binits{J.}},
\bauthor{\bsnm{Perona}, \binits{P.}},
\bauthor{\bsnm{Ramanan}, \binits{D.}},
\bauthor{\bsnm{Doll{\'a}r}, \binits{P.}},
\bauthor{\bsnm{Zitnick}, \binits{C.L.}}:
\bctitle{Microsoft coco: Common objects in context}.
In: \bbtitle{European Conference on Computer Vision},
pp. \bfpage{740}--\blpage{755}
(\byear{2014}).
\bcomment{Springer}
\end{bchapter}
\endbibitem

\bibitem{wikiart}
\begin{botherref}
\oauthor{\bsnm{Nichol}, \binits{K.}}:
Painter by numbers, wikiart
(2016)
\end{botherref}
\endbibitem

\bibitem{huang2018multimodal}
\begin{bchapter}
\bauthor{\bsnm{Huang}, \binits{X.}},
\bauthor{\bsnm{Liu}, \binits{M.-Y.}},
\bauthor{\bsnm{Belongie}, \binits{S.}},
\bauthor{\bsnm{Kautz}, \binits{J.}}:
\bctitle{Multimodal unsupervised image-to-image translation}.
In: \bbtitle{Proceedings of the European Conference on Computer Vision (ECCV)},
pp. \bfpage{172}--\blpage{189}
(\byear{2018})
\end{bchapter}
\endbibitem

\bibitem{yang2019unsupervised}
\begin{bchapter}
\bauthor{\bsnm{Yang}, \binits{J.}},
\bauthor{\bsnm{Dvornek}, \binits{N.C.}},
\bauthor{\bsnm{Zhang}, \binits{F.}},
\bauthor{\bsnm{Chapiro}, \binits{J.}},
\bauthor{\bsnm{Lin}, \binits{M.}},
\bauthor{\bsnm{Duncan}, \binits{J.S.}}:
\bctitle{Unsupervised domain adaptation via disentangled representations:
  Application to cross-modality liver segmentation}.
In: \bbtitle{International Conference on Medical Image Computing and
  Computer-Assisted Intervention},
pp. \bfpage{255}--\blpage{263}
(\byear{2019}).
\bcomment{Springer}
\end{bchapter}
\endbibitem

\bibitem{yang2020cross}
\begin{bchapter}
\bauthor{\bsnm{Yang}, \binits{J.}},
\bauthor{\bsnm{Li}, \binits{X.}},
\bauthor{\bsnm{Pak}, \binits{D.}},
\bauthor{\bsnm{Dvornek}, \binits{N.C.}},
\bauthor{\bsnm{Chapiro}, \binits{J.}},
\bauthor{\bsnm{Lin}, \binits{M.}},
\bauthor{\bsnm{Duncan}, \binits{J.S.}}:
\bctitle{Cross-modality segmentation by self-supervised semantic alignment in
  disentangled content space}.
In: \bbtitle{Domain Adaptation and Representation Transfer, and Distributed and
  Collaborative Learning},
pp. \bfpage{52}--\blpage{61}.
\bpublisher{Springer}, \blocation{???}
(\byear{2020})
\end{bchapter}
\endbibitem

\bibitem{zhang2018automatic}
\begin{botherref}
\oauthor{\bsnm{Zhang}, \binits{J.}},
\oauthor{\bsnm{Saha}, \binits{A.}},
\oauthor{\bsnm{Soher}, \binits{B.J.}},
\oauthor{\bsnm{Mazurowski}, \binits{M.A.}}:
Automatic deep learning-based normalization of breast dynamic contrast-enhanced
  magnetic resonance images.
arXiv preprint arXiv:1807.02152
(2018)
\end{botherref}
\endbibitem

\bibitem{yang2020mri}
\begin{barticle}
\bauthor{\bsnm{Yang}, \binits{Q.}},
\bauthor{\bsnm{Li}, \binits{N.}},
\bauthor{\bsnm{Zhao}, \binits{Z.}},
\bauthor{\bsnm{Fan}, \binits{X.}},
\bauthor{\bsnm{Eric}, \binits{I.}},
\bauthor{\bsnm{Chang}, \binits{C.}},
\bauthor{\bsnm{Xu}, \binits{Y.}}:
\batitle{Mri cross-modality image-to-image translation}.
\bjtitle{Scientific reports}
\bvolume{10}(\bissue{1}),
\bfpage{1}--\blpage{18}
(\byear{2020})
\end{barticle}
\endbibitem

\bibitem{mirza2014conditional}
\begin{botherref}
\oauthor{\bsnm{Mirza}, \binits{M.}},
\oauthor{\bsnm{Osindero}, \binits{S.}}:
Conditional generative adversarial nets.
arXiv preprint arXiv:1411.1784
(2014)
\end{botherref}
\endbibitem

\bibitem{modanwal2019normalization}
\begin{botherref}
\oauthor{\bsnm{Modanwal}, \binits{G.}},
\oauthor{\bsnm{Vellal}, \binits{A.}},
\oauthor{\bsnm{Mazurowski}, \binits{M.A.}}:
Normalization of breast mris using cycle-consistent generative adversarial
  networks.
arXiv preprint arXiv:1912.08061
(2019)
\end{botherref}
\endbibitem

\bibitem{zhu2017cyclegan}
\begin{bchapter}
\bauthor{\bsnm{Zhu}, \binits{J.-Y.}},
\bauthor{\bsnm{Park}, \binits{T.}},
\bauthor{\bsnm{Isola}, \binits{P.}},
\bauthor{\bsnm{Efros}, \binits{A.A.}}:
\bctitle{Unpaired image-to-image translation using cycle-consistent adversarial
  networks}.
In: \bbtitle{Proceedings of the IEEE International Conference on Computer
  Vision},
pp. \bfpage{2223}--\blpage{2232}
(\byear{2017})
\end{bchapter}
\endbibitem

\bibitem{ourdataset}
\begin{barticle}
\bauthor{\bsnm{Saha}, \binits{A.}},
\bauthor{\bsnm{Harowicz}, \binits{M.R.}},
\bauthor{\bsnm{Grimm}, \binits{L.J.}},
\bauthor{\bsnm{Kim}, \binits{C.E.}},
\bauthor{\bsnm{Ghate}, \binits{S.V.}},
\bauthor{\bsnm{Walsh}, \binits{R.}},
\bauthor{\bsnm{Mazurowski}, \binits{M.A.}}:
\batitle{A machine learning approach to radiogenomics of breast cancer: a study
  of 922 subjects and 529 dce-mri features}.
\bjtitle{British journal of cancer}
\bvolume{119}(\bissue{4}),
\bfpage{508}--\blpage{516}
(\byear{2018})
\end{barticle}
\endbibitem

\bibitem{transforms}
\begin{bchapter}
\bauthor{\bsnm{Zakhor}, \binits{A.}}:
\bctitle{Lecture 2. intensity transformation and spatial filtering}.
In: \bbtitle{EE225B: Digital Image Processing}
(\byear{2014}).
\bcomment{University of California, Berkeley}
\end{bchapter}
\endbibitem

\bibitem{ulyanov2016instance}
\begin{botherref}
\oauthor{\bsnm{Ulyanov}, \binits{D.}},
\oauthor{\bsnm{Vedaldi}, \binits{A.}},
\oauthor{\bsnm{Lempitsky}, \binits{V.}}:
Instance normalization: The missing ingredient for fast stylization.
arXiv preprint arXiv:1607.08022
(2016)
\end{botherref}
\endbibitem

\bibitem{kingma2014adam}
\begin{botherref}
\oauthor{\bsnm{Kingma}, \binits{D.P.}},
\oauthor{\bsnm{Ba}, \binits{J.}}:
Adam: A method for stochastic optimization.
arXiv preprint arXiv:1412.6980
(2014)
\end{botherref}
\endbibitem

\bibitem{he2015delving}
\begin{bchapter}
\bauthor{\bsnm{He}, \binits{K.}},
\bauthor{\bsnm{Zhang}, \binits{X.}},
\bauthor{\bsnm{Ren}, \binits{S.}},
\bauthor{\bsnm{Sun}, \binits{J.}}:
\bctitle{Delving deep into rectifiers: Surpassing human-level performance on
  imagenet classification}.
In: \bbtitle{Proceedings of the IEEE International Conference on Computer
  Vision},
pp. \bfpage{1026}--\blpage{1034}
(\byear{2015})
\end{bchapter}
\endbibitem

\bibitem{svc}
\begin{barticle}
\bauthor{\bsnm{Sch{\"o}lkopf}, \binits{B.}},
\bauthor{\bsnm{Smola}, \binits{A.J.}},
\bauthor{\bsnm{Williamson}, \binits{R.C.}},
\bauthor{\bsnm{Bartlett}, \binits{P.L.}}:
\batitle{New support vector algorithms}.
\bjtitle{Neural computation}
\bvolume{12}(\bissue{5}),
\bfpage{1207}--\blpage{1245}
(\byear{2000})
\end{barticle}
\endbibitem

\end{thebibliography}


\clearpage
\begin{appendices}
\section*{Supplementary Material for ``Deep Learning for Breast MRI Style Transfer with Limited Training Data''}
\section{Data}
\label{app:data}
In this section, we will give a more detailed description of our dataset and the process of its creation, beyond that of Section \ref{sec:data}. 

In this study, we experimented with the Breast Cancer DCE-MR (Dynamic Contrast-Enhanced Magnetic Resonance) dataset of \cite{ourdataset}, which includes MRI scans of 922 breast cancer patients over the course of a decade. From this, we include GE Healthcare MRI scans from 628 subjects, and each scan volume consists of more than 160 physically adjacent 2D axial image slices. To focus on the salient parts of the volumes and to avoid redundancies in our dataset, for each volume we selected a single slice from the middle 50\% of each patient volume used in our study to be used in our final dataset.

All images have a $512\times512$ pixel resolution, and were pre-processed by assigning the top 1\% of pixel values in the entire dataset to a value of 255, followed by linearly scaling the remaining pixel intensities to the 0-255 range. The data was randomly divided at the patient level, and 528 datapoints were used to produce the training set. Out of the remaining images, 50 were kept as a test set, and the other 50 were used for validation. 25 images from a Siemens scanner were created the same way, to be used in Section \ref{sec:exp:classifyscanners}.

This data is publicly available on the Cancer Imaging Archive (TCIA), at \url{https://doi.org/10.7937/TCIA.e3sv-re93}.

\section{Image Transformation Functions/Training Styles}
\label{app:transforms}
Here we give the explicit formulae for the parametrically-randomized image transformation functions used to simulate style in training. Note that after a randomized transformation function is applied to a training image, the image is normalized to fall within the pixel range of 0-255. 
\paragraph{The Linear Transformation}
The simplest intensity transfer function/transformation is linear with respect to input pixel intensity $I_\mathrm{in}$ with some slope $m_\mathrm{lin}$ (contrast change) and intercept $b_\mathrm{lin}$ (brightness offset), with output pixel intensity $I_\mathrm{out}$. We ``randomize'' this function during training with
\begin{equation}
    I_\mathrm{out} = T_\mathrm{lin}(I_\mathrm{in}) = m_\mathrm{lin}I_\mathrm{in} + b_\mathrm{lin},
\end{equation}
where given the continuous uniform distribution $\mathcal{U}(\cdot,\cdot)$, $m_\mathrm{lin}=\tan\theta_\mathrm{lin}$, $\theta_\mathrm{lin}\sim\mathcal{U}({\pi/8},{3\pi/8})$ and $b_\mathrm{lin}\sim\mathcal{U}({-20},{20})$ are sampled at each new iteration of training when the linear transformation is chosen.

\paragraph{The Negative Transformation}
The negative of an image is produced by subtracting each pixel from the maximum possible intensity value $I_\mathrm{max}$, e.g. for an 8-bit image $I_\mathrm{max} = 2^8 - 1 = 255$. As such, the transformation function $T_\mathrm{neg}$ used to produce an image negative can be written pixel-wise as $I_\mathrm{out} = T_\mathrm{neg}(I_\mathrm{in}) = I_\mathrm{max} - I_\mathrm{in}$, where $I_\mathrm{max}$ is the maximum pixel intensity value for the image, and $I_\mathrm{out}$ and $I_\mathrm{in}\geq 0$ are the output and input pixel intensity values, respectively. We ``randomize'' this function during training with
\begin{equation}
    I_\mathrm{out} = T_\mathrm{neg}(I_\mathrm{in}) = m_\mathrm{neg}I_\mathrm{in} + b_\mathrm{neg},
\end{equation}
with $m_\mathrm{neg}=\tan\theta_\mathrm{neg}$, $\theta_\mathrm{neg}\sim\mathcal{U}({-3\pi/8},{-\pi/8})$ and $b_\mathrm{neg}\sim\mathcal{U}({235},{275})$.

\paragraph{The Log Transformation}
The log (logarithm) transformation of an image can be expressed as a pixel-wise transformation function $T_\mathrm{log}$ as $I_\mathrm{out} = T_\mathrm{log}(I_\mathrm{in}) = c_\mathrm{log}\log(1+I_\mathrm{in})$,
where $c_\mathrm{log}$ is a scaling constant given by $c_\mathrm{log}=I_\mathrm{max}/[\log (1 + I_\mathrm{max, img})]$, with $I_\mathrm{max, img}$ being the maximum pixel value in the given image. This  factor is chosen to ensure that the range of output pixel values does not exceed $I_\mathrm{max}$. In practice the log transformation is used to map a narrow band of low-intensity input values to a wide range of output intensities. We randomize this transformation via
\begin{equation}
\label{eq:trans_log}
I_\mathrm{out} = T_\mathrm{log}(I_\mathrm{in}) = \tilde{c}_\mathrm{log}\log(1+I_\mathrm{in}),
\end{equation}
where $\tilde{c}_\mathrm{log} = ac_\mathrm{log}$ with $a\sim\mathcal{U}({0.7},{1.3})$.

\paragraph{The Power-Law (Gamma) Transformation}
The power-law (gamma) transformation can be mathematically expressed as a pixel-wise transformation function with $I_\mathrm{out} = T_\mathrm{pow}(I_\mathrm{in}) = c\left({I_\mathrm{in}}/{I_\mathrm{max}}\right)^\gamma$,
where similar to $c_\mathrm{log}$, $c_\mathrm{pow}=I_\mathrm{max}$ is a scaling constant chosen to give the correct range of possible $I_\mathrm{out}$ values. The exponential parameter $\gamma$ can be any chosen as any $\gamma>0$. This transformation is used for gamma correction, a method important for the correct displaying of images on digital screens. This transformation is randomized with the scheme
\begin{equation}
\label{eq:trans_gamma}
I_\mathrm{out} = T_\mathrm{pow}(I_\mathrm{in}) = c\left(\frac{I_\mathrm{in}}{I_\mathrm{max}}\right)^{\tilde{\gamma}} ,
\end{equation}
where $\tilde{\gamma}=2^\alpha$ with $\alpha\sim\mathcal{U}({-5},{5})$.

\paragraph{The Piecewise-Linear Transformation}
Piecewise-linear transformation functions are constructed by conjoining different linear functions that have disjoint intensity ranges. A common application of these transformation functions is for \textit{contrast stretching}, a method for expanding an image's intensity values to span the full possible range of intensities. Here, we use a three segment piecewise-linear transformation function defined ``randomly'' according to 
\begin{equation}
\begin{split}
I_\mathrm{out} &= T_\mathrm{pw}(I_\mathrm{in}) \\
&= 
\begin{cases} 
\vspace{0.75em}\displaystyle\frac{s_1}{r_1}I_\mathrm{in} &\mathrm{if} \, I_\mathrm{in}\in [0, r_1] \\
\vspace{0.75em}\displaystyle\frac{s_2 - s_1}{r_2-r_1}(I_\mathrm{in}-r_1) + s_1 &\mathrm{if} \, I_\mathrm{in}\in [r_1, r_2] \\
\displaystyle\frac{I_\mathrm{max} - s_2}{I_\mathrm{max}-r_2}(I_\mathrm{in}-r_2) + s_2 &\mathrm{if} \, I_\mathrm{in}\in [r_2, I_\mathrm{max}],
\end{cases}
\end{split}
\end{equation}
where $r_1 < r_2, s_1 < s_2$ are parameters chosen by the scheme
\begin{equation}
\begin{split}
    & \qquad r_1\sim\mathcal{U}({55},{95}) \qquad r_2\sim\mathcal{U}({130},{170})\\
    & \qquad s_1\sim\mathcal{U}({35},{75}) \qquad s_2\sim\mathcal{U}({205},{245}).
\end{split}
\end{equation}

\paragraph{The Sobel X and Y Transformations}
\label{sec:trans_sobel}
The Sobel X and Y transformations, or operators, are edge-detection image transformations defined by applying $3\times 3$ convolutional kernels to the input image. The X and Y kernel matrices are defined to approximate the derivatives that quantify horizontal and vertical changes of input pixel intensity values, respectively. The X and Y operators are as such defined respectively with
\begin{equation}
\begin{split}
    &\textbf{I}_{\mathrm{out}, x} = 
    \begin{bmatrix}
    1 & 0 & -1 \\
    2 & 0 & -2 \\
    1 & 0 & -1
    \end{bmatrix} * \textbf{I}_\mathrm{in}
    \quad\mathrm{and}\quad\\
    &\textbf{I}_{\mathrm{out}, y} = 
    \begin{bmatrix}
    1 & 2 & 1 \\
    0 & 0 & 0 \\
    -1 & -2 & -1
    \end{bmatrix} * \textbf{I}_\mathrm{in},
\end{split}
\end{equation}
where $*$ indicates the linear convolution operation, $\textbf{I}_\mathrm{in}$ is the input image defined as a matrix of pixel intensity values, and $\textbf{I}_{\mathrm{out}, x}, \textbf{I}_{\mathrm{out}, y}$ are the output images of the X and Y transformations, respectively.

\section{Full Expansion of Cross-Domain Reconstruction Triplet Loss (Equation \eqref{eq:loss_cross})}
\label{app:fullcrossloss}
\begin{equation}
\begin{split}
\mathcal{L}_{\mathrm{cross}} &= 
\mathbb{E}\left[\left\Vert G\left(E^c(X_2), E^s(X_1)\right)-X_2\right\Vert_1\right] \\
&+ \mathbb{E}\left[\left\Vert G\left(E^c(X_1), E^s(X_2)\right)-X_1\right\Vert_1\right] \\
&+ \mathbb{E}\left[\left\Vert G\left(E^c(X_2), E^s(T(X_1))\right)-T(X_2)\right\Vert_1\right] \\
&+ \mathbb{E}\left[\left\Vert G\left(E^c(X_1), E^s(T(X_2))\right)-T(X_1)\right\Vert_1\right] \\
&+ \mathbb{E}\left[\left\Vert G\left(E^c(X_2), E^s(T(X_2))\right)-T(X_2)\right\Vert_1\right] \\
&+ \mathbb{E}\left[\left\Vert G\left(E^c(X_1), E^s(T(X_1))\right)-T(X_1)\right\Vert_1\right] \\
&+ \mathbb{E}\left[\left\Vert G\left(E^c(T(X_2)), E^s(X_1)\right)-X_2\right\Vert_1\right] \\
&+ \mathbb{E}\left[\left\Vert G\left(E^c(T(X_1)), E^s(X_2)\right)-X_1\right\Vert_1\right] \\
&+ \mathbb{E}\left[\left\Vert G\left(E^c(T(X_2)), E^s(X_2)\right)-X_2\right\Vert_1\right] \\
&+ \mathbb{E}\left[\left\Vert G\left(E^c(T(X_1)), E^s(X_1)\right)-X_1\right\Vert_1\right] \\
&+ \mathbb{E}\left[\left\Vert G\left(E^c(T(X_2)), E^s(T(X_1))\right)-T(X_2)\right\Vert_1\right] \\
&+ \mathbb{E}\left[\left\Vert G\left(E^c(T(X_1)), E^s(T(X_2))\right)-T(X_1)\right\Vert_1\right]. \\
\end{split}
\end{equation}

\section{Exploring Style Encoding Mechanics}
\label{app:stylecodes}

In the previous two sections we explored two different applications of StyleMapper. In this section, we will take a different approach, and explore how style encoding can differ between styles seen and unseen in training, and between images of different styles and of the same style. To begin, in this section we will use ``fixed'' versions of the randomized training style transformations $T_i$ described in Section \ref{sec:transforms}, for purposes of consistency and comparability of results. In particular, the randomized parameters will be fixed to the means of their respective sampling distributions: the linear function will be fixed to identity, via $m_\mathrm{lin}=1, b_\mathrm{lin}=0$. The negative transformation will be fixed to $m_\mathrm{neg}=-1, b_\mathrm{neg}=255=I_\mathrm{max}$; log: $a=1$; power-law/gamma: $\tilde{\gamma}=0.5$; piecewise-linear: $r_1=75, r_2=150, s_1=55, s_2=225$. The Sobel X and Y transformations are unchanged.

We will begin by comparing the style codes that the trained style encoder extracts from images of different styles. In particular, we take a test set $\{X_{\mathrm{target}}\}$ of 25 raw DCE-MR images, which we can then apply one of the aforementioned transformations $T_i$ to, in order to obtain the corresponding set of transformed images $\{T_i(X_{\mathrm{target}})\}$. Finally, from here we can input each of these transformed images to the trained style encoder $E^s$ to obtain the set of style codes for each these images, $\{s^{T_i}_{\mathrm{target}}=E^s(T_i(X_{\mathrm{target}}))\}$. 

\begin{figure}[htbp]
    \centering
    \includegraphics[width=1.0\linewidth]{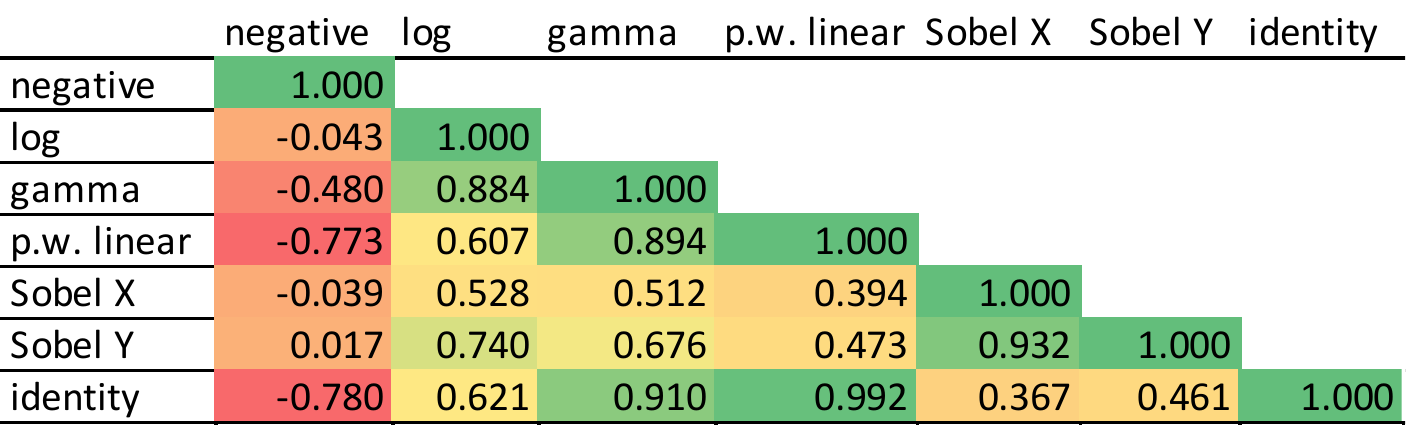}
    \caption{\textbf{Comparing style codes extracted for different styles.} Entries are cosine-similarities (Equation \eqref{eq:sim}) of style codes extracted from pairs of images of the same content but different styles, i.e. two different image transformations applied to the same test image, averaged over the entire test set. The rows and columns specify which pair of transformations/styles that the encoded images originated from. This figure is recommended to be viewed in color.}
    \label{fig:differentstyles}
\end{figure}

Now, we can estimate the similarity of two style code vectors $s_1,s_2\in\mathbb{R}^8$ with a cosine-similarity/normalized inner product
\begin{equation}
    \label{eq:sim}
    \mathrm{sim}(s_1,s_2)=\frac{s_1^Ts_2}{||s_1||_2||s_2||_2},
\end{equation}
where $\mathrm{sim}(s_1,s_2)\in[-1,1] \,\,\forall s_1, s_2$. In order to examine how the style encoder encodes various styles differently, we compare codes extracted from images of different styles but, with the same content. We do this by computing $\mathrm{sim}(s^{T_i}_{\text{target,$k$}},s^{T_j}_{\text{target,$k$}})$ for each image $X_{\text{target,$k$}}$ in the first half of the test set ($k=1,\ldots,25$), where the pair of style codes $s^{T_i}_{\text{target,$k$}}, s^{T_j}_{\text{target,$k$}}$ are obtained from applying each pair of different transformations $T_i,T_j$ to $X_{\text{target,$k$}}$. We then average the similarities over all of the images in the test set for each pair of transformations; the results of this are shown in Figure \ref{fig:differentstyles}.

As we saw in Section \ref{sec:exp:oneshot}, the ability of a trained style encoder to learn the style code of a given target style is mostly independent of the number of images $N_\mathrm{target}$ of this target style seen by the encoder; we can see this numerically by observing the distribution of style codes obtained from applying a style encoder on a distribution of images of a certain single style, as follows. In particular, for the set of style codes extracted from images transformed with one of the transformations $T_i$, $\{s^{T_i}_{\mathrm{target}}\}$, we can estimate how similar style codes corresponding to this transformation/style are to each other, by averaging over the similarities between all possible pairs of different style codes from $\{s^{T_i}_{\mathrm{target}}\}$. 

These results are given for each of the examined styles in Table \ref{tab:samestyle}. On average, style codes of different images of the same style barely differ, indicated by the average similarities all being very close to unity, and the small standard deviations thereof. In other words, the style encoder is very consistent with what code it assigns to a particular style. This is true even for styles not seen in training: for a style encoder trained on all transformations \textit{except} for the power-law transformation, the style codes extracted from power-law-style test images have an average paired similarity of $0.985$ with standard deviation $0.023$, very close to the corresponding result for the model trained on power-law styles in Table \ref{tab:samestyle}.

\begin{table}[]
    \caption{Average and standard deviation of cosine similarity (Equation \eqref{eq:sim}) of all pairs of style codes extracted from images of the same style.}
    \label{tab:samestyle}
    \centering
    \begin{tabular}{llll}
        \toprule
        Style & Avg. similarity & std. deviation \\
        \midrule
        Negative & 0.995 & 0.006 \\
        Log & 0.974 & 0.037 \\
        Gamma/power-law & 0.984 & 0.023 \\
        Piecewise-linear & 0.979 & 0.028 \\
        Sobel X & 0.989 & 0.020 \\
        Sobel Y & 0.990 & 0.016 \\
        Identity & 0.990 & 0.013 \\
        \bottomrule
    \end{tabular}
\end{table}

\end{appendices}

\end{document}